\documentstyle[aps,prl,multicol,epsf]{revtex}

\begin{document}
\begin{title}
  {\bf Switching between different vortex states in 2-dimensional
    easy-plane magnets due to an ac magnetic field}
\end{title}

\author{Yuri Gaididei}
\address{ 
Institute for Theoretical Physics, 252143 Kiev, Ukraine}
\author{Till Kamppeter and Franz G. Mertens}
\address{
Physikalisches Institut, Universit\"at Bayreuth, D-95440 Bayreuth,
Germany}

\author{A. R. Bishop}
\address{ 
Theoretical Division and Center for Nonlinear Studies, Los Alamos
National Laboratory, MS B262, Los Alamos, New Mexico 87545}

\maketitle

\bigskip 
\begin{abstract}
Using a discrete model of 2-dimensional easy-plane classical
ferromagnets, we propose that a rotating magnetic field in the easy
plane can switch a vortex from one polarization to the opposite one if
the amplitude exceeds a threshold value, but the backward process does
not occur. Such switches are indeed observed
in computer simulations.
\end{abstract}


\begin{multicols}{2}


There is a growing interest in non-equilibrium dynamics of
quasi-two-dimensional magnetic materials
\cite{jongh,gav,pom,head}. Many magnetic properties of these materials
are well-described by the classical two-dimensional Heisenberg model with
easy-plane symmetry.
In this model vortices play a very important role. They cause a
topological phase  transition \cite{berezinskii}, 
and they contribute to   the so-called
central peaks in inelastic neutron scattering experiments  
\cite{Regnault87,Mikeska91,Wiesler94}
that  arise from  the translational motion of vortices \cite{fgm87,gouvea}.

There are two types of static vortex solutions,
 depending on the anisotropy strength $\delta$ (see below) 
of the Heisenberg exchange
interaction \cite{gouvea}:
in-plane vortices  for which all spins lie in the easy
$xy$-plane, and 
out-of-plane vortices which  exhibits a localized structure of
the $z$-components of the spins around the vortex center. In addition
to the vorticity $q = \pm1, \, \pm 2, \ldots$, the out-of-plane vortices
have a second topological charge $p$. This is denoted as  ''polarization''
because its sign determines the side of the $xy-$plane to which the
out-of-plane vortex structure points.

The aim of the present Letter is to investigate the internal dynamic
of the vortices in the discrete two-dimensional Heisenberg ferromagnet
in the presence of an ac magnetic field
in the easy-plane.  
It is shown that  switching between vortex states with different 
polarization occurs if the amplitude of the
field is larger than a threshold value.
   
The easy-plane Hamiltonian  for classical spins 
$\vec{S}_{\vec{n}}=
S\{\sin\theta_{\vec{n}}\cos\Phi_{\vec{n}},
\sin\theta_{\vec{n}}\sin\Phi_{\vec{n}},
\cos\theta_{\vec{n}}\}$ 
  located on sites $\vec{n}=(n_x,n_y)$ of a quadratic 
  lattice has the form
\begin{eqnarray}
\label{ham}
H=-J \,\sum_{\vec{n},\vec{a}} \{(1-\delta)\,M_{\vec{n}}\,
M_{\vec{n}-\vec{a}}+\nonumber\\
\sqrt{1-M^2_{\vec{n}}}\,\sqrt{1-M^2_{\vec{n}-\vec{a}}}\,
\cos(\Phi_{\vec{n}}-\Phi_{\vec{n}-\vec{a}})\}
\end{eqnarray}
where $\delta$ is the anisotropy parameter ($0<\delta\leq\,1$), 
$ J > 0$ is the exchange constant  and $\vec{a}$
 is  the vector which connects a site with nearest 
neighbors.
$S^{z}_{\vec{n}}\equiv M_{\vec{n}}=\cos\theta_{\vec{n}}$ is 
the on-site magnetization. In what follows we set $J=1$ and 
$S=1$.

It is  known \cite{gouvea,wysin94} that 
for $\delta\,>\,\delta_c$ where the critical value 
of the anisotropy
parameter $\delta_c$ depends on the lattice type(e.g. 
for square lattices $\delta_c \simeq\,0.3$) the in-plane
vortex with $M_{\vec{n}}=0$ and the azimuthal angles $\Phi_{\vec{n}}$ 
satisfying the equation
$\sum_{\vec{a}}\sin(\Phi^0_{\vec{n}}-
\Phi^0_{\vec{n}-\vec{a}})=0$,
is stable. For $\delta\,<\,\delta_c$ the 
in-plane vortex becomes unstable and  an out-of-plane vortex is created.
It exhibits a localized structure of the $M_{\vec{n}}$ components
around the vortex center. In the case of a circular system of the radius $L$ 
with free
boundary conditions the azimuthal angles $\Phi_{\vec{n}}$  for both types
of vortices are
approximately given by
\begin{eqnarray}
\label{az}
\Phi_{\vec{n}}=q\arctan\left(\frac{n_y-Y}{n_x-X}\right)-
q\arctan\left(\frac{n_y-\bar{Y}}{n_x-\bar{X}}\right),
\end{eqnarray}
where a constant phase has been omitted.  $X$ and $Y$ are the
coordinates of the vortex center, and $\bar{X}=X L^2/R^2$, $\bar{Y}=Y
L^2/R^2$ ($R^2 = X^2+Y^2$) are the coordinates of the "image"-vortex.
In the case of fixed boundary conditions the sign in front of the
second term in Eq. (\ref{az}) is reversed.

We are interested here in the vortex dynamics under the influence of a
 spatially uniform
in-plane ac magnetic field $\vec{h}(t)=h (\cos\omega t,\sin\omega t,0)$.

 The interaction
of the field with the spin system has the form
\begin{eqnarray}
\label{int}
V(t)=-h\,\sum_{\vec{n}}\sqrt{1-M^2_{\vec{n}}}\,\cos (\Phi_{\vec{n}}-\omega t).
\end{eqnarray} 
The spin dynamics is described by the Landau-Lifshitz equation

\begin{eqnarray}
\label{lln}
\dot{\Phi}_{\vec{n}}=\frac{\partial }{\partial M_{\vec{n}}}\left(H+V(t)\right)-
\frac{\gamma}{1-M^2_{\vec{n}}}\frac{\partial H}{\partial
\Phi_{\vec{n}}}
\nonumber\\
\dot{M}_{\vec{n}}=-\frac{\partial }{\partial 
\Phi_{\vec{n}}}\left(H+V(t)\right)-
\gamma\,(1-M^2_{\vec{n}})\,\frac{\partial H}{\partial M_{\vec{n}}}.
\end{eqnarray}
The last   terms in  Eqs (\ref{lln})
  represent  damping  \cite{iida63}.

To clarify the behavior of out-of-plane vortices in the
presence of the ac field,  we have
numerically integrated the Landau-Lifshitz equation (\ref{lln}) for a
large square lattice in which we cut out a circle with radius $L=24$
using both free and fixed boundary conditions with
$\delta=0.1,\gamma=0.002$. We used relatively weak
ac fields so as not to  change the ground state significantly. 
The integration time was 12,000 time units with 
time step 0.01. First we used an
out-of-plane vortex with  polarization p=1 as the initial
condition and a  clockwise rotating magnetic field with the
frequency $\omega=-0.1$: This  is close to the frequency of
the lowest radially symmetric eigenmode  in the presence of a
vortex \cite{wysin94,wysin95,ivanov}. We observed that for all $h \leq
h_{cr}= 0.0025$ the vortex with p=1 remains  the stable configuration
but for  
 $h\,>\,h_{cr}$ a flip to the state with the opposite polarization
($p=-1$) occurs ( Fig 1 ). Using the same initial condition but
changing the direction of rotation  of the magnetic field
($\omega=0.1$) we  observed the switching only  when $h > 0.02$. But in
contrast to the previous case when the vortex after switching had a
well-defined  core structure, now the out-of-plane structure is almost
completely destroyed by spin waves.

Another set of simulations was performed using a static out-of-plane
vortex state with the same $h$  but  polarization $p=-1$ as initial
condition. 
We
found that the flip  occurs only for a counter-clockwise rotating magnetic
field $\omega >0$. The results of extensive simulations
may be summarized as follows for both types of boundary conditions:
\begin{itemize}
\item Flips between oppositely polarized  states  take place
 under the action of the ac magnetic field
  when  $h > h_{cr}$ ( Fig. 2).
\item The threshold value $h_{cr}$ depends on the product $\omega\,p$
  and    not  on the
  vorticity. The threshold value in the case when the polarization vector is
  anti-parallel to the angular velocity vector
  $\vec{\omega}=(0,0,\omega)$ is  much smaller than when these vectors 
are parallel. 
\item Flips  are uni-directional. When  $\omega p\,< \,0$,  the final
  state is characterized by a well-defined core structure, while for
  $\omega p\,> \,0$ the core structure is destroyed.
\end{itemize}

The basic reason for the switching can be easily
understood by using 
 the  frame of reference  which rotates together with the  magnetic
field. In this frame  there exists an inertial force
 equivalent to a magnetic field aligned along the angular
velocity $\vec{\omega}$. Then the vortex states with different
polarization are nonequivalent and switching processes become
energetically favored. This  cannot explain, however,
why the threshold of the switching is a non-monotonic function  of the
frequency  $\omega$ ( Fig. 2). 
To gain deeper insight 
 we  need in a reduced form of the Hamiltonian (\ref{ham})
which  takes into account  both types of vortices: in-plane
and out-of-plane. As  a topological charge,$p$, is 
 conserved in the continuum limit only.
 Thus  the switching between states with different
 polarization  is due to  lattice discreteness. 

We consider
the near-critical case $|(\delta-\delta_c)/(1-\delta_c)|\ll 1$
 when the out-of-plane spin deviations $M_{\vec{n}}$  are small, and
 assume also  smooth dependence  of the  deviations 
$~~\phi_{\vec{n}}=
\Phi_{\vec{n}}-\Phi^0_{\vec{n}}~~$ from the static vortex structure on
 the spatial variable $\vec{n}$. 
In this case, applying the transformation
\begin{eqnarray}
\label{trans}
M_{\vec{n}}=\sum_{\nu}{\cal L}_{\vec{n}\,\nu}m_{\nu},~~~ 
\phi_{\vec{n}}=\sum_{\nu}{\cal K }_{\vec{n}\,\nu}\psi_{\nu}
\end{eqnarray}
where the coefficients ${\cal L}_{\vec{n},\nu},~{\cal
  K}_{\vec{n},\nu}~$ satisfy the set of equations ${\cal
  L}_{\vec{n},\nu}=2 \sum_{\vec{a}}\left({\cal K}_{\vec{n},\nu}- {\cal
    K}_{\vec{n}-\vec{a},\nu}\right)
\cos(\Phi^0_{\vec{n}}-\Phi^0_{\vec{n}-\vec{a}}), \\
\mu_\nu {\cal
  K}_{\vec{n},\nu}=2\!\sum_{\vec{a}}\!\left(-(1-\delta){\cal
    L}_{\vec{n}-\vec{a},\nu} +{\cal
    L}_{\vec{n},\nu}\cos(\Phi^0_{\vec{n}}-\Phi^0_{\vec{n}-\vec{a}})\right)$
we transform the harmonic part of the Hamiltonian (\ref{ham}) obtained
in the vicinity of the static in-plane vortex,
\begin{eqnarray}
\label{efh}
H_0=\frac{1}{2}\,\,\sum_{\vec{n},\vec{a}}(\phi_{\vec{n}}-
\phi_{\vec{n}-\vec{a}})^2\cos(\Phi^0_{\vec{n}}-\Phi^0_{\vec{n}-\vec{a}})
-\nonumber\\
 \,\sum_{\vec{n},\vec{a}}\left(\lambda\,M_{\vec{n}}\,
M_{\vec{n}-\vec{a}}-M^2_{\vec{n}}\,
\cos(\Phi^0_{\vec{n}}-\Phi^0_{\vec{n}-\vec{a}})\right)
\end{eqnarray}
to the principal axis coordinates \newline
$H_0=\frac{1}{2}\,\sum_{\nu}\left(\psi^2_\nu+\mu_{\nu}\,m^2_{\nu}\right)$.

In Refs \cite{wysin95,ivanov} the linear
eigenmodes of the easy-plane ferromagnet with  Hamiltonian  
(\ref{ham}) in the presence of a vortex were investigated. 
The lowest radially
symmetric mode, which is localized near the  vortex center and
describes  the in-phase motion of the core spins,  
becomes soft when $\delta$ approaches  $\delta_c$. 
In other words, this mode is responsible for the in-plane vortex
instability. 

The corresponding eigenvalue, say $\mu_1$, 
 becomes 
negative when $\delta < \delta_c$: $~~\mu_1=B \, (\delta-\delta_c)~$ with 
$B$  a numerical coefficient.  Inserting 
the transformation (\ref{trans}) into the Hamiltonian $H_1=H-H_{0}$ and
keeping only 
the soft  eigenmode
$\nu=1$,  we obtain an effective soft-mode Hamiltonian
\begin{eqnarray}
\label{effh}
H_s= \frac{1}{2} (\psi^2_1+\mu_1\, m^2_1)+\frac{A}{4} \, m^4_1,
\end{eqnarray} 
where   terms $m^2_1 \psi^2_1$
and $\psi^4_1$  which are  unimportant in the near-critical case and
all higher order terms have been  omitted. 
$A=\frac{1}{2}\, \sum_{\vec{n},\vec{a}}\left({\cal L}^2_{\vec{n},1}-
{\cal L}^2_{\vec{n}-\vec{a},1}\right)^2\,\cos (\Phi^0_{\vec{n}}-
\Phi^0_{\vec{n}-\vec{a}})$  is a positive constant.

We see from Eqs (\ref{int}) 
and (\ref{az}) that a spatially uniform in-plane magnetic field cannot
excite the radially symmetric soft mode when the vortex is situated at
the center of the system. Switching
 can occur 
only as a result of nonlinear mixing between the radially symmetric mode and 
non-symmetric vortex  modes which do interact with the spatially uniform 
alternating external field. This may take place in  large 
systems where the motion of the vortex is frozen \cite{nik}.
However, in a relatively small,finite  system with free boundary 
conditions the vortex is attracted by its image and moves along an 
unwinding spiral trajectory (see e.g. \cite{volk})
towards the boundary.  
Our numerical experiments do show that switching events  in
general occur only
when vortex is at a finite distance from the center. The
vortex  center motion is very slow (with a frequency $ \sim 1/L^2$).
So we can consider the switching process
with fixed vortex position, say $X=R\cos \chi,Y=R\sin \chi$. Inserting
Eqs (\ref{trans}) and (\ref{az}) into Eq. (\ref{int}) and assuming that
the vortex is far from the boundaries ($R\ll L$),  we find that the 
effective interaction of the in-plane ac magnetic field
with the soft mode:  
\begin{eqnarray} 
\label{inte}
V_s(t)&=& h \left(a_1\,\psi_1 \sin (\omega t)-
h \frac{1}{2}
  (a_2\,\psi^2_1 + b\,m^2_1 )\,\cos(\omega t)\right) \nonumber \\
\end{eqnarray}
where $a_l=\sum_{\vec{n}}\frac{R \sqrt{n_x^2+n_y^2}}{2 L^2}
\,\left({\cal K} _{\vec{n},1}\right)^l,~(l=1,2),$ and
$~~b=\sum_{\vec{n}}\frac{R \sqrt{n_x^2+n_y^2}}{2 L^2} \,\left({\cal L}
  _{\vec{n},1}\right)^2$.  An effective interaction of the same form
can be obtained by taking into account the fact that the vortex
structure is velocity dependent \cite{gouvea} and symmetric about the
direction of the vortex motion. The constant phase shift ($q\chi$)
plays no essential role and was omitted.

From Eqs (\ref{effh}) and (\ref{inte}) we find that in the soft mode
approach the dynamics  is governed by
\begin{eqnarray}
\label{core}
\dot{m_1}&=&-\psi_1-h \left(a_1\sin\omega t - a_2 \psi_1 \cos\omega
  t\right)\nonumber\\
  &&\qquad\qquad-\gamma (\mu_1 m_1 + A m^3_1),\nonumber\\
\dot{\psi_1}&=&\mu_1  m_1+A m^3_1-\gamma \psi_1-h b m_1 \cos\omega t.
\end{eqnarray}
Here the nonlinear terms  $m_1^n, (n > 3)$ and $m_1^2  \psi_1$ have been
neglected. An example of the core dynamics based on Eqs (\ref{core})
is presented in Fig. 3. These results are in good
agreement with those from the numerical integration of the full
 Landau-Lifshitz equations (\ref{lln}).

To clarify the physical meaning it is
 convenient to write the set of equations (\ref{core}) as a single
equation for $m_1$. Near the threshold  ($~|\mu_1|\ll 1$),
in the limit of small damping ( $\gamma \ll 1$),  and for a not too
strong amplitude of the external magnetic field ($~a_2 h \leq 1$)  we
obtain from Eqs (\ref{core}) an effective equation for $m_1$  
 $\ddot{m_1}+ \gamma\, \dot{m_1}+ \mu_1\, m_1+A\, m^3_1+h (a_1\, \omega -b\,
m_1)\cos\omega t=0$.
Thus  the vortex core dynamics  is
analogous to the dynamics of a particle  in a 
double-well potential under the action of direct  and parametric
forces.In order to  estimate
$h_{cr}$ we use a heuristic approach proposed by Moon \cite{moon}.
Here the switching occurs when the
 particle reaches the maximum velocity on the homoclinic
orbit. Considering the particle motion in the potential well centered
at $m_1=p\,\sqrt{\mu_1}$  and using  
multiple-scale analysis  for small $\gamma$,  $h$ and $|\omega^2-\omega^2_0|\ll
\,1$,  where $\omega_0=\sqrt{2|\mu_1|}$ is the frequency of harmonic
oscillations near the bottom of the well, we find $m_1\approx
 \sqrt{|\mu_1|}\left(p+M(\omega)\cos(\omega t)\right)$, with
\begin{eqnarray} 
\label{est}
M^2\,\bigg(\left(\omega^2_0-\omega^2-\frac{3}{2}\omega^2_0\,M^2\right)^2+
\gamma^2\,\omega^2\bigg)=\nonumber\\
2 A \frac{h^2}{\omega^4_0}\,(a_1\,\omega-b \sqrt{|\mu_1|} \,p)^2.
\end{eqnarray}
The maximum velocity on the homoclinic orbit is $\frac{|\mu_1|}{4
  A}$.The switching occurs
  when
$\omega M(\omega)=\alpha\frac{|\mu_1|}{4
  A}$,  where $\alpha\approx 1$ is an empirical parameter.From 
 Eq. (\ref{est}) we find $ h_{cr}(\omega)$  in the form
\begin{eqnarray}
\label{crit}
 &&\frac{\alpha \omega^2_0}{\sqrt{2
    A}}\,\frac{1}{|\Omega \left(a_1\sqrt{2}\Omega-b
    \,p\right)|}\,\sqrt{\left(1-\Omega^2-
\frac{3}{8}\frac{\alpha^2}{\Omega^2}\right)^2+
\frac{\gamma^2}{\omega^2_0}\,\Omega^2}\nonumber \\
\end{eqnarray}  
where $\Omega=\omega/\omega_0$. 
The condition (\ref{crit}) is in  agreement with the results of
numerical simulations: 
the function $h_{cr}(\omega)$ has  minima near the  frequency
of the soft mode $\omega\approx \pm \omega_0$. It  is highly
asymmetric ( Fig. 4).
For small amplitudes of the counter-clockwise rotating 
magnetic field ($\omega >0$), 
the switching condition $h\,>\,h_{cr}$ can be fulfilled
only for the vortex with initial polarization $p=-1$,  while for a
clockwise rotating field  the condition can be fulfilled for the
vortex with opposite polarization $p=1$. Qualitatively the same
results may be obtained by using the Melnikov function approach \cite{mel}.

In conclusion, we have shown that the polarization of out-of- plane vortices
in easy-plane ferromagnets can be changed  by applying an ac
magnetic field. Flips occur more
easily when the polarization of the
vortex is anti-parallel to the angular velocity 
$\vec{\omega}$.  It can take place only in 
discrete systems, and they are uni-directional events.

\section{Acknowledgements}
Yu. Gaididei is grateful for the hospitality of the University of Bayreuth 
and the Los Alamos National Laboratory where this work was
performed.Work at Los Alamos is performed under auspices of the
U.S.D.o.E.
 Discussions with Darryl Holm  and Gary Wysin are acknowledged.

\subsection*{Figures}

\centerline{\epsfxsize=8.5truecm \epsffile{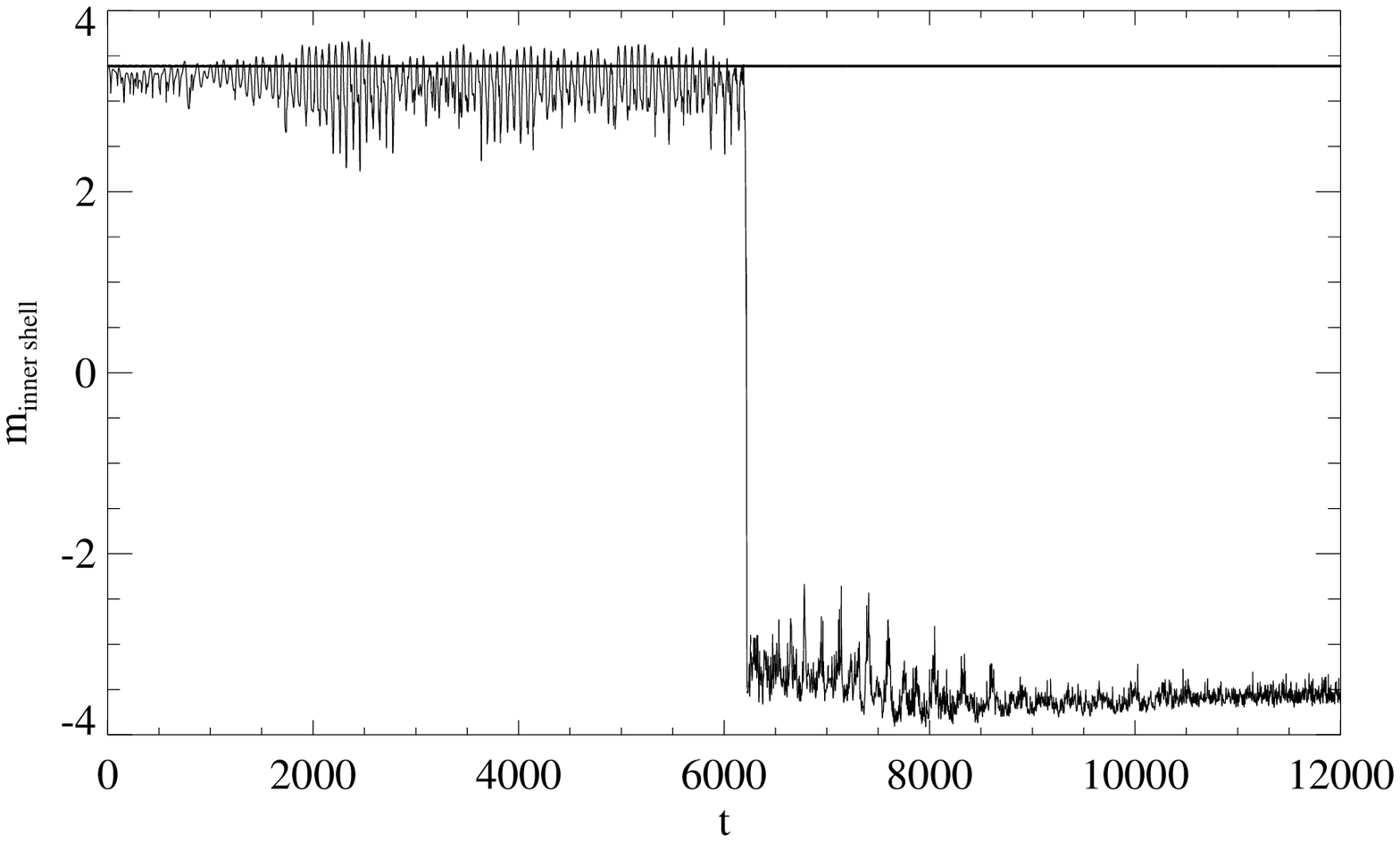}}

{\bf Figure 1:} Switching from the  state with positive polarization
to the state with negative polarization of the out-of-plane vortex due
to
a clockwise rotating magnetic field with the frequency
$\omega=-0.1$. The damping
constant $\gamma=0.002$. The lower curve shows the time evolution of the
magnetization of the  inner shell when the field amplitude 
$h=3\times 10^{-3}$ is  above the threshold value,the straight 
line corresponds to  $h= 10^{-3}$ (the amplitude of oscillations in 
this case is very small $\approx 0.005$ and therefore they are not
seen in the figure)
 
\centerline{\epsfxsize=8.5truecm \epsffile{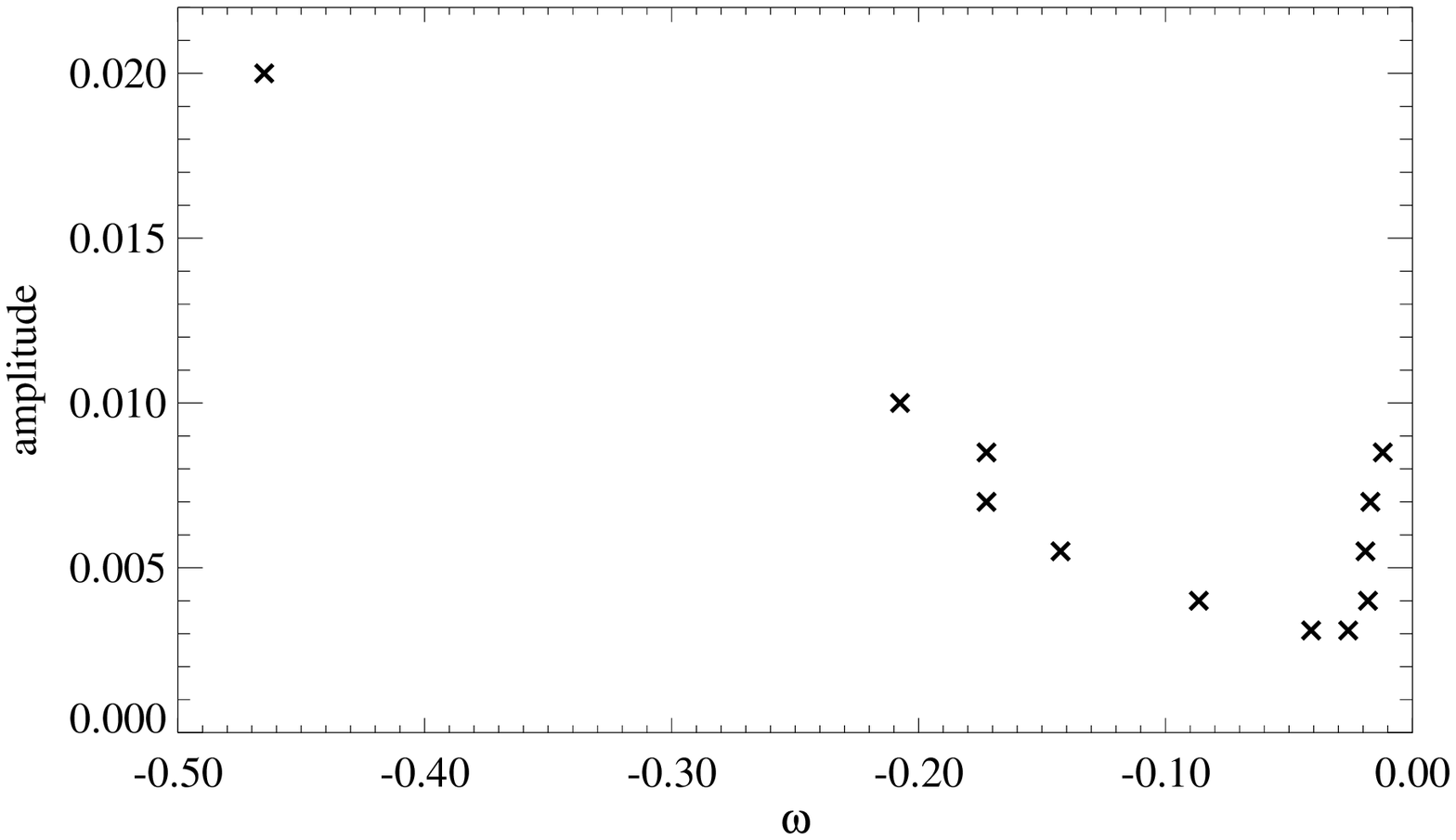}}

{\bf Figure 2:} Threshold value of the field amplitude $h_{cr}$ versus
$\omega$ obtained from the numerical integration of the full 
Landau-Lifshitz equations.
                        
\centerline{\epsfxsize=8.5truecm \epsffile{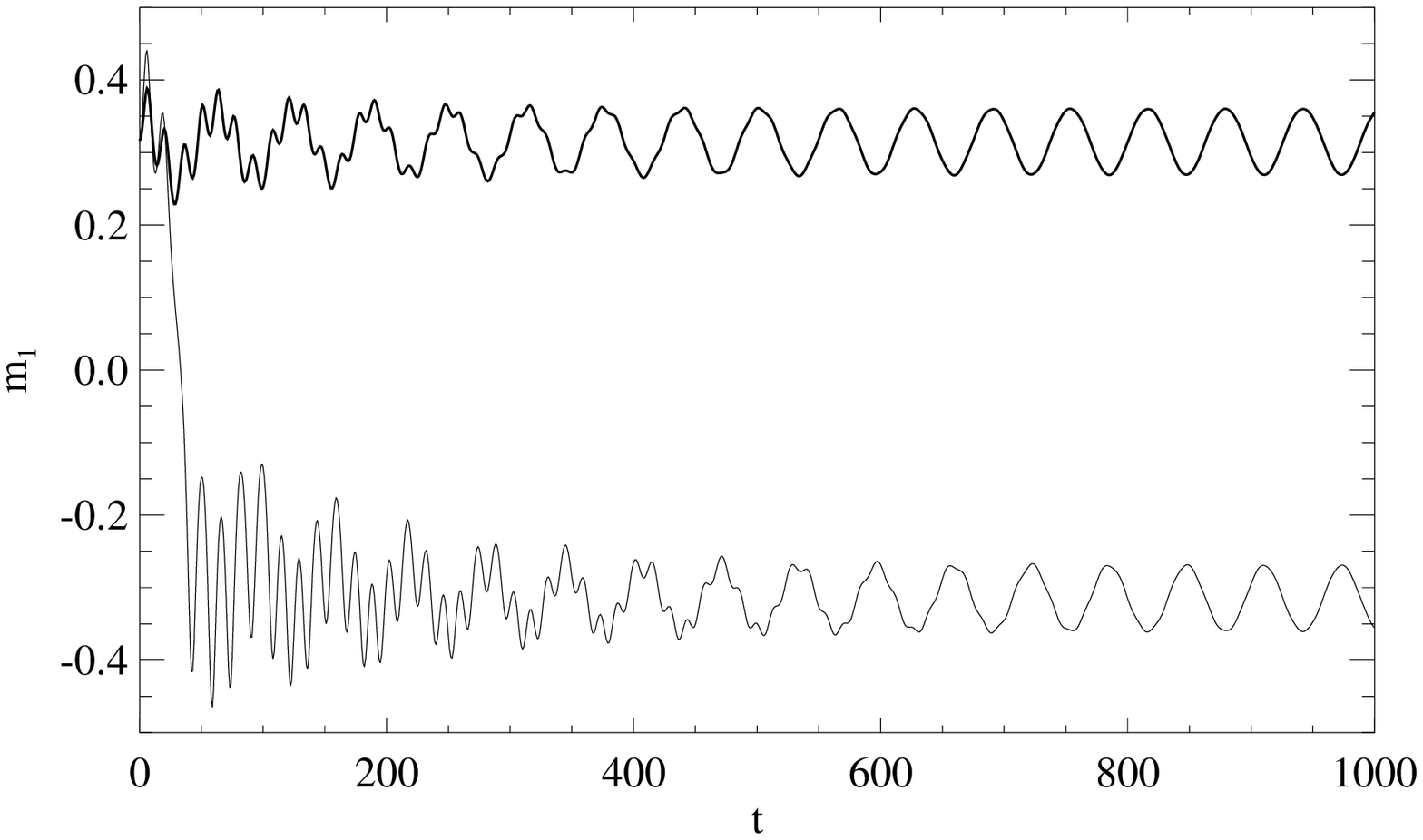}}

{\bf Figure 3:} Time evolution of the vortex core magnetization 
in the presence  of the clockwise ( upper curve) and
counter-clockwise (lower curve) 
rotating magnetic field based on Eqs (\ref{core}). The parameters are
 $A=a_1=a_2=b=1,\mu_1=-0.1,\omega=\pm 0.1,h=0.04,\gamma=0.01$. Initial
polarization $p=+1$.

\centerline{\epsfxsize=8.5truecm \epsffile{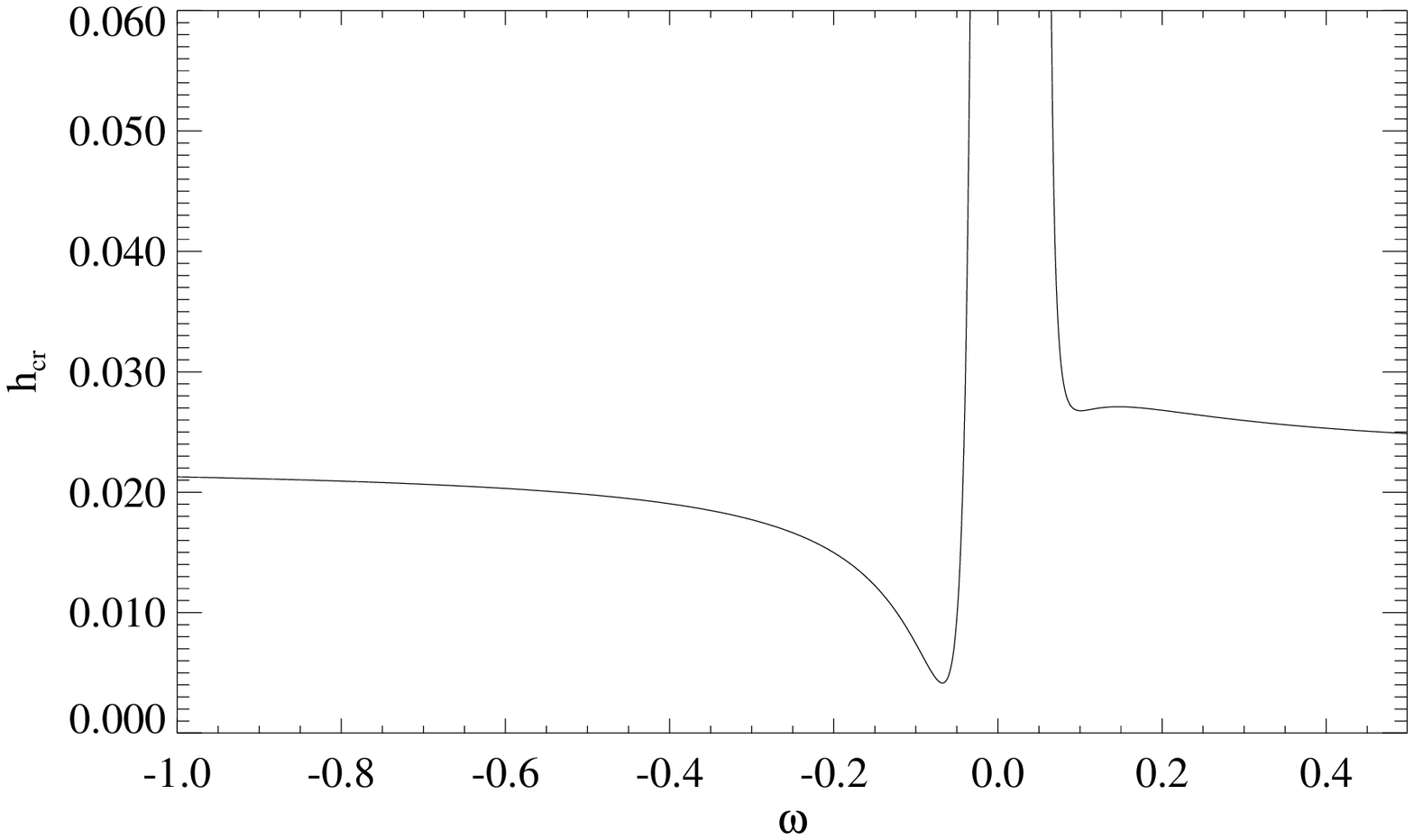}}

{\bf Figure 4:}  Threshold value of the field amplitude $h_{cr}$ versus
$\omega$, as  given by Eq. (\ref{crit}). The parameters
used are $a_1=b, A\,b'/\alpha=0.01,\omega_0=0.08,\gamma=0.002$.           


\end{multicols}



\begin{thebibliography}{99}

\bibitem{jongh} L. J. de Jongh, in {\it Magnetic Properties of Layered
    Transition Metal Compounds},  ed.
  by L. J. de Jongh (Kluwer Academic Press, Dortrecht, 1990).
\bibitem{gav} P. Gaveau, J. P. Boucher, L. P. Regnault, J. Y. Henry,
    J. Appl. Phys. {\bf 69}, 6228 (1991).
\bibitem{pom} M. Pomerantz, Surf. Sci. {\bf 142},556 (1984).
\bibitem{head} D. I. Head, B. H. Blott, and D. Melville,
  J. Phys. (Paris) Colloq. {\bf 49}, C8-164 (1988).
\bibitem{berezinskii}  V. Berezenskii, Sov. Phys. JETP {\bf 32}, 493
  (1970); J. M. Kosterliz, and D. J. Thouless, J. Phys. C{\bf 6},1181
  (1973).
\bibitem{Regnault87} L. P. Regnault and J. Rossat-Mignod, in {\it
    Magnetic Properties of Layered Transition Metal Compounds}, ed.
  by L. J. de Jongh (Kluwer Academic Press, Dortrecht, 1990).
  
\bibitem{Mikeska91} H.-J. Mikeska and M. Steiner,  Adv. Phys. {\bf
    40},191 (1991).

 \bibitem{Wiesler94} D. G. Wiesler, H. Zabel, and S. M. Shapiro, Z.
  Phys. B {\bf 93}, 277 (1994)
  
\bibitem{fgm87} F. G. Mertens, A. R.  Bishop, G. M. Wysin, and C.
  Kawabata, Phys. Rev. Lett. {\bf 59}, 117 (1987); 
Phys. Rev. B {\bf 39}, 591 (1989)
  
\bibitem{gouvea} M. E. Gouvea, G. M. Wysin, A. R. Bishop, and F. G. Mertens,
Phys. Rev. B {\bf 39},11840 (1989).
\bibitem{wysin94} G. M. Wysin, Phys. Rev. B {\bf 49}, 
8780 (1994)
\bibitem {iida63} S. Iida, J. Phys. Chem. Solids {\bf 24}, 625 (1963).
  

\bibitem{wysin95} G. M. Wysin, A. R. V\"o lkel, Phys. Rev. B {\bf 52}, 
7412 (1995).
\bibitem{ivanov} B. A. Ivanov, H. J. Schnitzer, F. G. Mertens, and
  G. M. Wysin, Phys. Rev. B {\bf58},8464 (1998).

\bibitem{ivanov96} B. A. Ivanov, A. K. Kolezhuk, and G. M. Wysin,
  Phys. Rev. Lett. {\bf 76}, 511 (1996)



\bibitem{nik} A. V. Nikiforov and E. B. Sonin, Sov. Phys. JETP {\bf
58},373 (1983).

\bibitem{volk}A. R. V{\"o}lkel, G. M. Wysin, F. G. Mertens,
  A. R. Bishop, and
  H. J. Schnitzer, Phys.Rev. B {\bf 50},12711 (1994).
\bibitem{moon} F. C. Moon, { \it Chaotic Vibrations} (Wiley, New York, 1987).
\bibitem{mel} V. K. Melnikov, Trans. Moskow Math. Soc. {\bf12},1
(1963).
\end{thebibliography}
\end{document}